\documentclass[a4paper, 12pt]{scrartcl}

\usepackage[breaklinks=true,pdfborder={0 0 0}]{hyperref}

\usepackage[utf8]{inputenc}
\usepackage[T1]{fontenc}
\usepackage[english]{babel}
\usepackage{csquotes}

\usepackage[section]{placeins}
\usepackage{graphicx}
\usepackage{subcaption}

\usepackage{siunitx}

\usepackage[shortcuts]{extdash}

\usepackage{amsmath}
\usepackage{amsfonts}
\usepackage{amssymb}

\usepackage[backend=biber,style=numeric,autopunct, sorting=none, alldates=year, block=ragged, autolang=hyphen, isbn=false]{biblatex}

\DeclareGraphicsExtensions{.pdf,.png,.jpg}

\title{Development of a GEM based TPC Readout for ILD}
\author{Paul Malek\thanks{DESY Hamburg}}
\date{\large Talk presented at the International Workshop on Future Linear Colliders (LCWS2016), Morioka, Japan, 5-9 December 2016. C16-12-05.4.}

\begin{document}

\maketitle

\begin{abstract}
For the International Large Detector (ILD), foreseen to be built at the International Linear Collider (ILC), a Time Projection Chamber (TPC) is intended to be used as the main tracking detector.
The amplification will be provided by Micro Pattern Gaseous Detectors (MPGDs).
One option is the use of Gas Electron Multipliers (GEM) in combination with a segmented pad readout plane.
The TPC group at DESY developed a modular system implementing a triple GEM stack mounted on thin ceramic grids.
This material choice allows for high mechanical rigidity of the support structure at a reduced amount of material and dead area compared to commonly used GRP frames.

This contribution gives an overview of the current status of this system. This includes a discussion of points we wanted to improve over the last generation of modules and what was implemented in the newest version.
Also improvements in the production process of the modules, which ensure a consistent quality and present a step towards possible batch production.
\end{abstract}

\section{Project Background}
\label{sec:background}


This development project is part of the design effort for the \emph{International Large Detector} (ILD) at the planned \emph{International Linear Collider} (ILC) \cite{ILCSummary,ILD_TDR_detectors}.
The FLC-TPC group at DESY Hamburg is part of the LCTPC collaboration, which is driving the efforts of developing a TPC for the ILD.
Different readout and amplification technologies based on \emph{Micro Pattern Gaseous Detectors} (MPGD), are currently studied within the collaboration.
FLC-TPC committed to designing a self supporting readout composed of GEMs \cite{SAULI1997531} mounted on thin ceramic grids \cite{steder13}.

\section{The DESY GridGEM Module}
\label{sec:module}

\begin{figure}[thp]
  \centering
  \hfill
  \begin{subfigure}{0.45\textwidth}
    \centering
    \includegraphics[width=\textwidth,height=0.25\textheight,keepaspectratio=true]{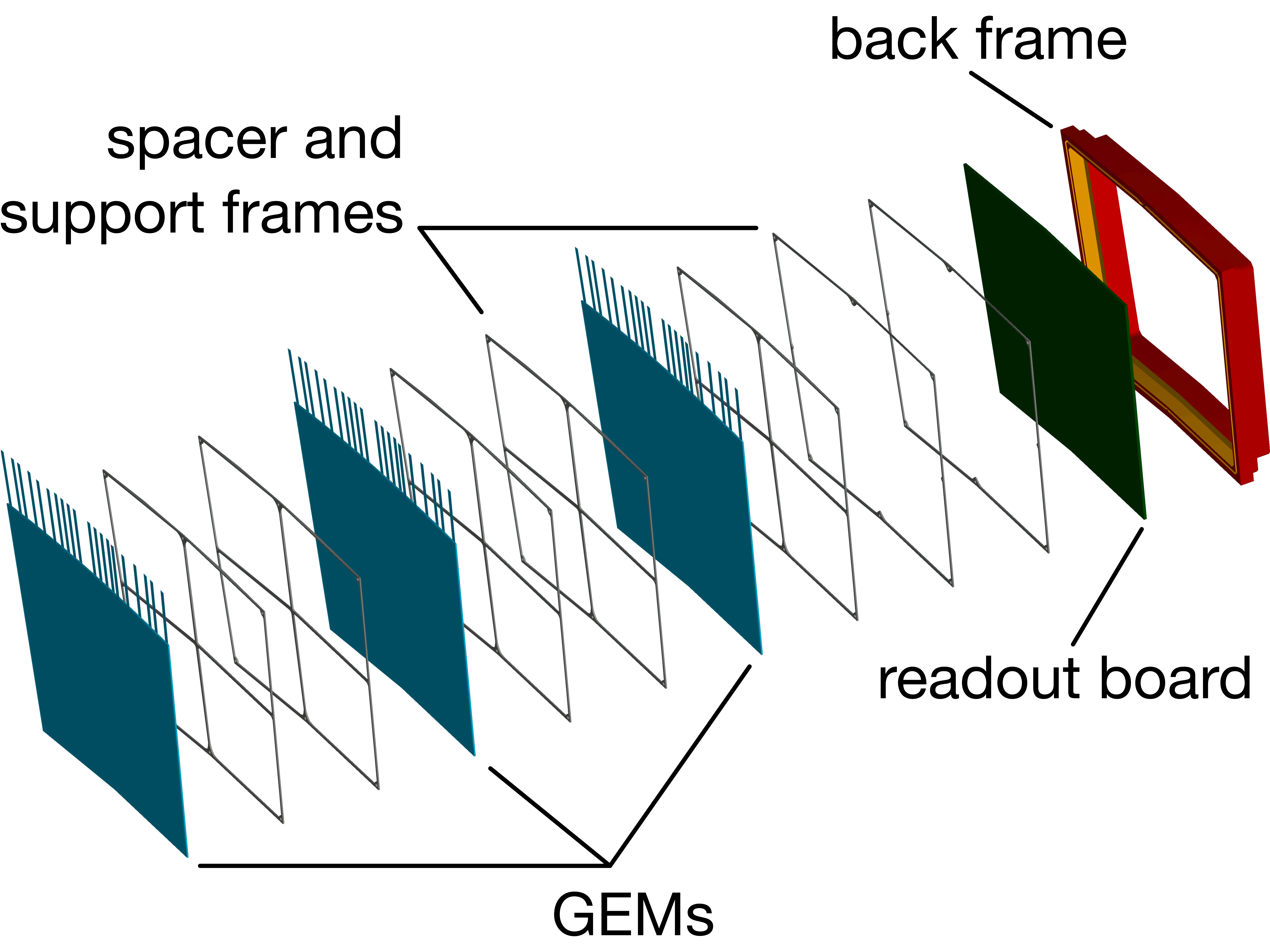}
    \caption{Explosion view of the readout module \cite{lctpc16}.\label{fig:module_explosion}}
  \end{subfigure}
  \hfill
  \begin{subfigure}{0.45\textwidth}
    \centering
    \includegraphics[width=\textwidth,height=0.25\textheight,keepaspectratio=true]{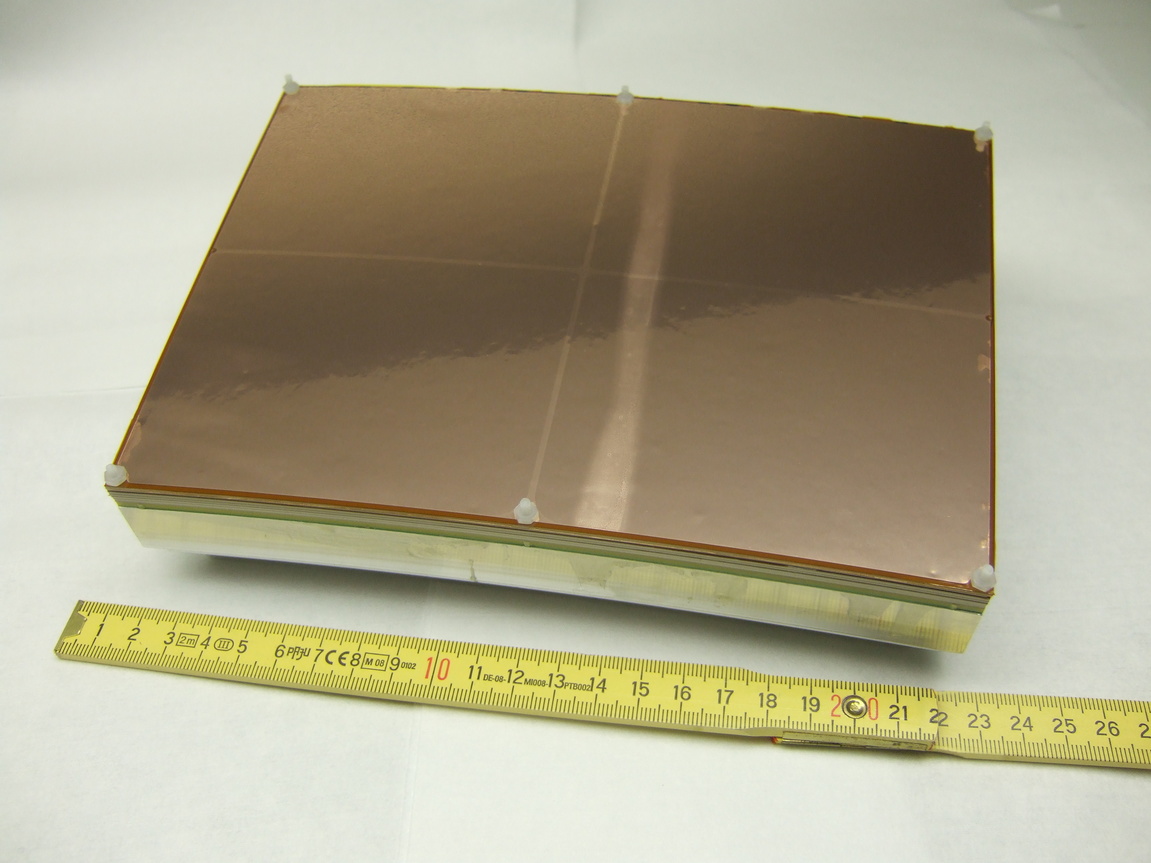}
    \caption{Picture of an assembled GEM module.\label{fig:module}}
  \end{subfigure}
  \hfill
  \caption{The DESY GridGEM module.}
\end{figure}

The mechanical base of the module is an aluminium back frame onto which a PCB containing the readout-pad plane is glued.
This structure is used to mount the module in the TPC end plate and provides the gas tightness.
On top of the pad board, the amplification structure is build up out of three GEMs mounted on \SI{1}{mm} high alumina-ceramic frames.
These frames also serve as spacers between the GEM foils to define transfer and induction gaps as seen in \autoref{fig:module_explosion}.
A fully assembled module is shown in \autoref{fig:module}.
Since the frame bars are only \SI{1.4}{mm} wide, the modules have an active area of about \SI{95}{\percent}.

The feasibility of using GEMs mounted on thin ceramic grids in a TPC amplification stage was first shown with \SI{10x10}{cm} GEMs \cite{steder13}.
In the next step, the point resolution was studied in the DESY II test beam with a series of full-size, \SI{17x22}{cm}, modules and it was shown in \cite{mueller16} that the design goals could be reached.
Still, there are some performance parameters to test (i.e. dE/dx, double hit resolution) and technical issues to address.

\subsection{Possible Points of Improvement}
\label{sec:improvements}
For a new iteration of the module the flatness of the mounted GEMs was improved to increase the homogeneity of the electric field and the gas gain.
To this end, a tool for the merging of grids and GEMs was developed to ensure a consistently high flatness.
This is described in \autoref{sec:stretchingtool}.

\begin{figure}[thp]
  \centering
  \includegraphics[width=0.9\textwidth,height=0.25\textheight,scale=1,keepaspectratio=true]{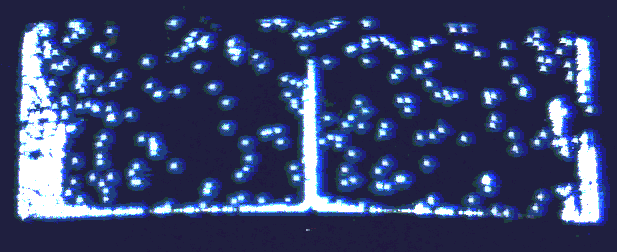}
  \caption{Image of discharges on two GEM sectors accumulated over time. Along the frames an increased number of discharges was observed.}
  \label{fig:discharges}
\end{figure}

In a different study, the GEMs were tested under extreme high voltage settings, far beyond normal operating conditions.
Here, a higher accumulation of discharges was observed along the edges of framed GEMs (\autoref{fig:discharges}).
This was found to be correlated to glue from the mounting process spilling onto the active area of the GEM.
The glue choice and application parameters were further optimized to achieve a continuous glue bond between GEM and grid with a minimum amount of glue.
In addition, to be on the safe side, the margin between the grid and the active GEM area was slightly increased.

\begin{figure}[thp]
  \centering
  \hfill
  \begin{subfigure}{0.45\textwidth}
    \centering
    \includegraphics[clip=true, trim=6.75cm 3.8cm 4.5cm 3.8cm, width=\textwidth, height=0.3\textheight, keepaspectratio=true]{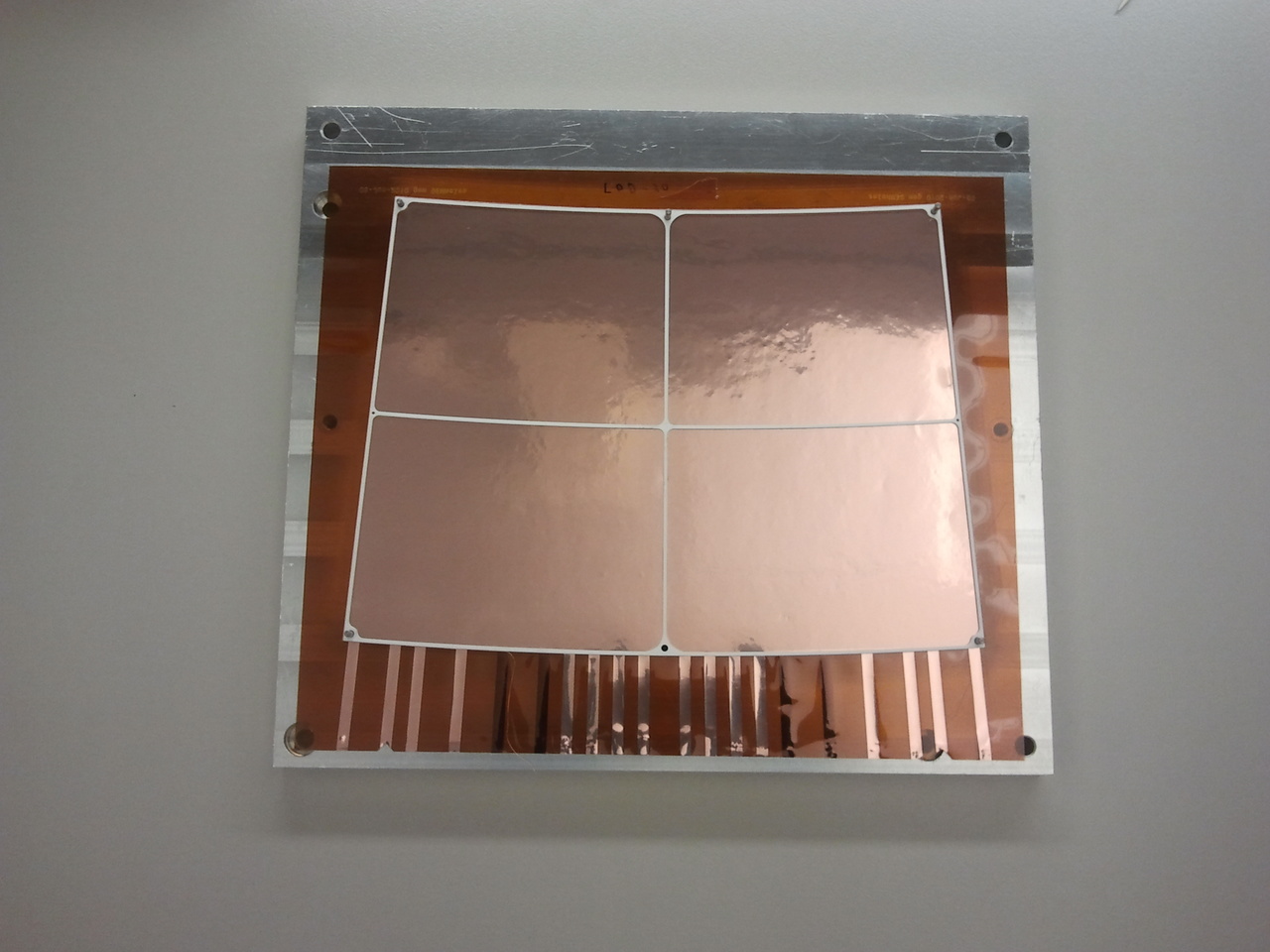}
    \caption{Last generation GEM\label{fig:oldgem}}
  \end{subfigure}
  \hfill
  \begin{subfigure}{0.45\textwidth}
    \centering
    \includegraphics[clip=true, trim=10.5cm 1.1cm 8cm 2.2cm, width=\textwidth, height=0.3\textheight, keepaspectratio=true]{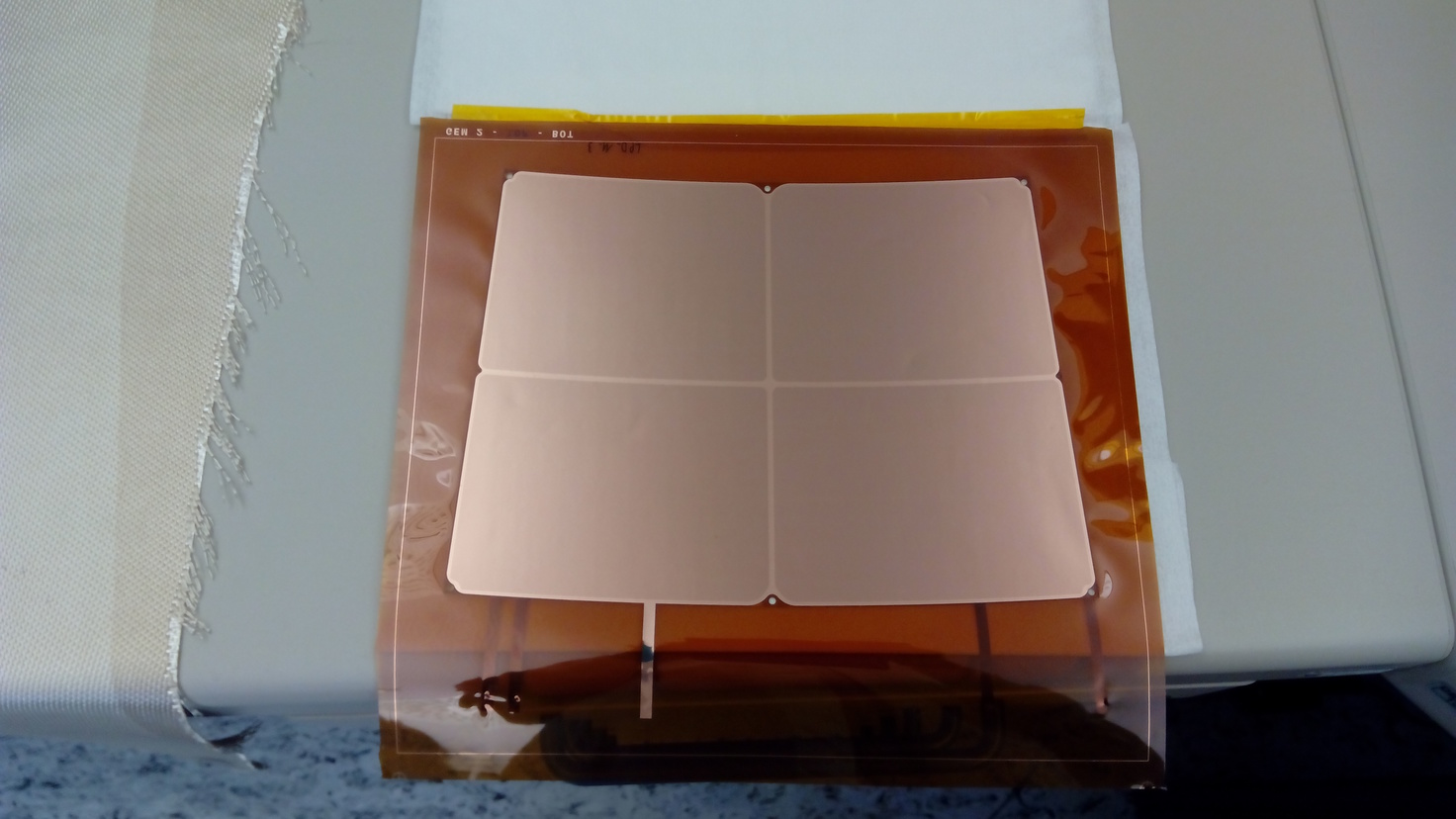}
    \caption{Current generation GEM\label{fig:newgem}}
  \end{subfigure}
  \hfill
  \caption{\subref*{fig:oldgem}) Picture of a GEM of the last generation with a ceramic frame on top. The HV connection lines are visible at the bottom of the image. \subref*{fig:newgem}) Image of a new GEM with only the needed number of HV connections.\label{fig:gems}}
\end{figure}

In the last generation of modules, the GEMs were all produced from the same design.
This meant that in all GEMs the high voltage (HV) connection lines for all positions in the stack were present (\autoref{fig:oldgem}).
When cutting the additional lines sharp copper edges were left, which potentially could be a source of discharges.
To avoid this problem the new GEMs were produced with only the needed HV lines for each layer (\autoref{fig:newgem}).

\section{Influence of GEM Flatness on the TPC Performance}
\label{sec:performance}

\subsection{Impact on the dE/dx Measurement}
\label{sec:dEdx}
\begin{figure}[thp]
  \centering
  \hfill
  \begin{subfigure}{0.5\textwidth}
    \centering
    \includegraphics[width=\textwidth, height=0.25\textheight, keepaspectratio=true]{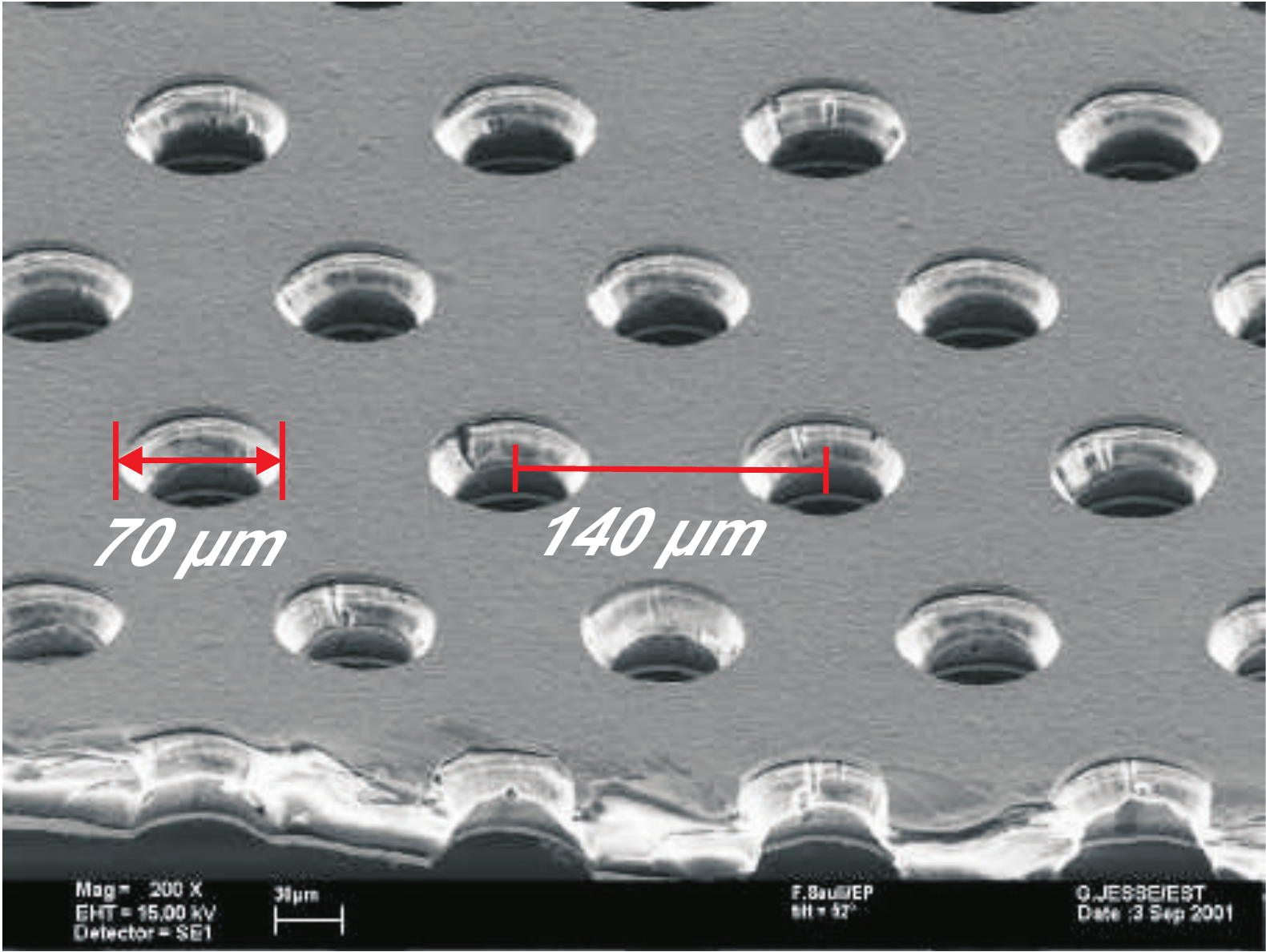}
    \caption{Microscope picture of a GEM produced at the CERN workshop \cite{gdd14}.\label{fig:gemmicro}}
  \end{subfigure}
  \hfill
  \begin{subfigure}{0.4\textwidth}
    \centering
    \includegraphics[width=\textwidth, height=0.25\textheight, keepaspectratio=true]{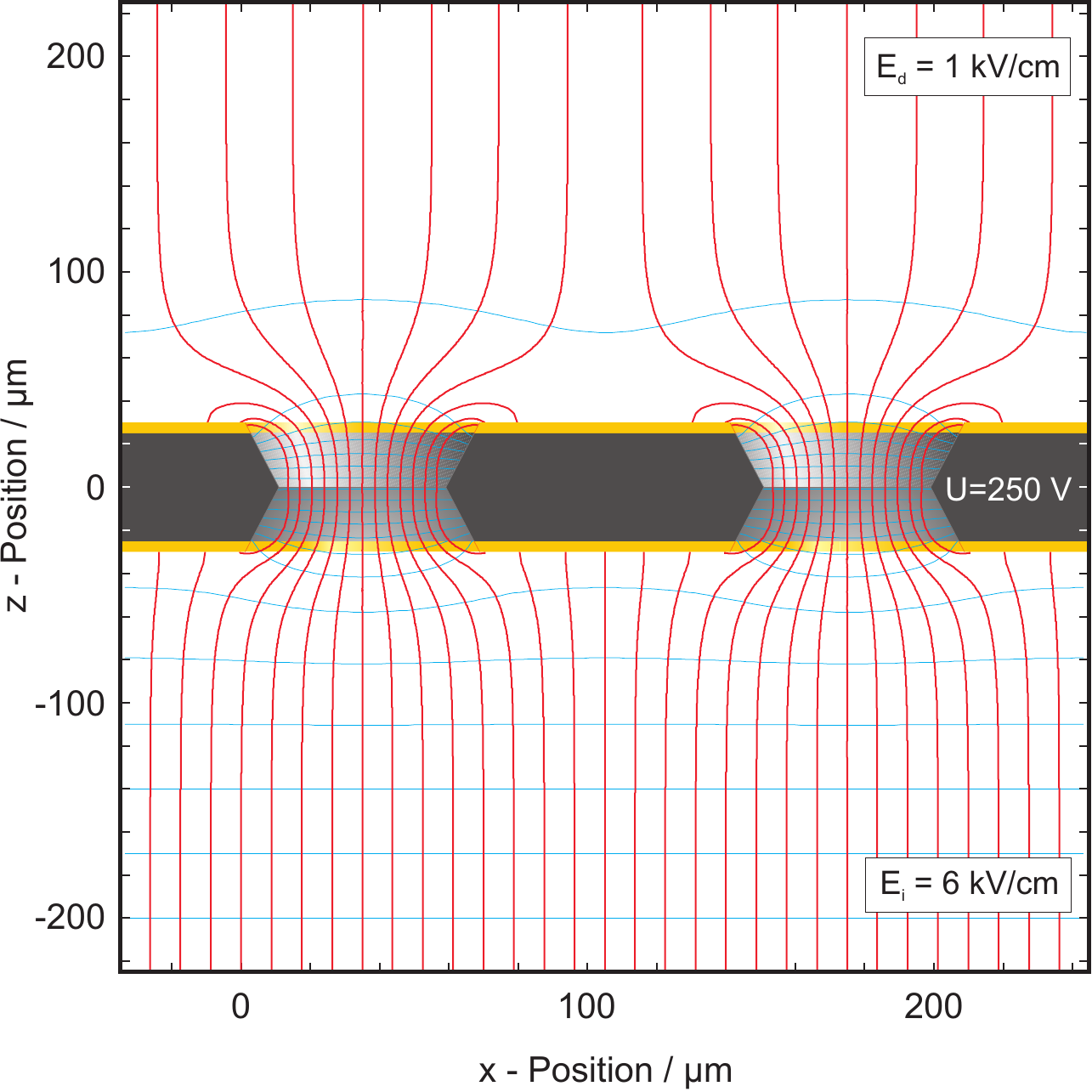}
    \caption{Field lines of the electrical field near a GEM \cite{sobloher02,schaefer06}.\label{fig:gemfields}}
  \end{subfigure}
  \hfill
  \caption{Working principle of GEMs\label{fig:gem_principle}}
\end{figure}

\begin{figure}[thp]
  \centering
  \includegraphics[clip=true, trim=0 0 0 4cm, width=0.9\textwidth, height=0.25\textheight, keepaspectratio=true]{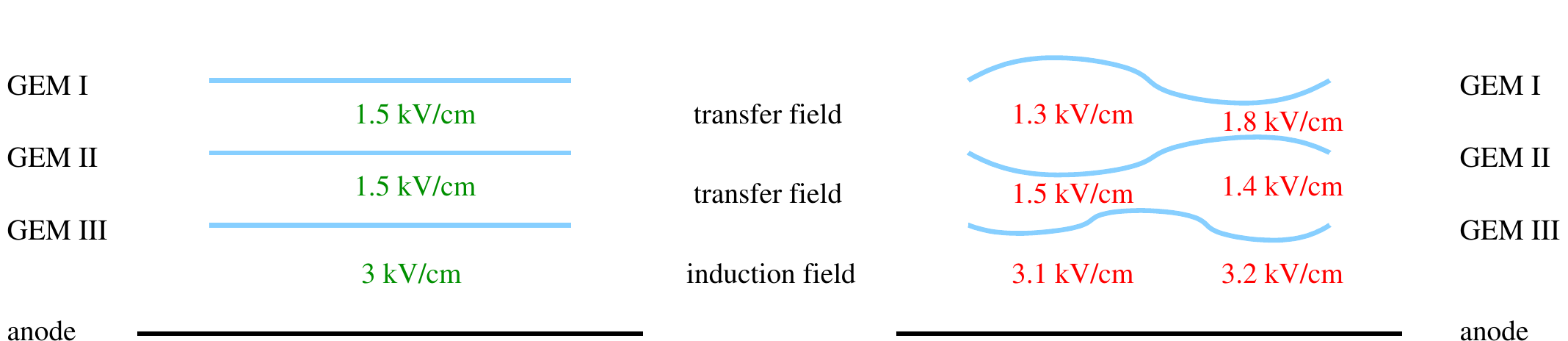}
  \caption{Example of a nominal field configuration in the GEM stack and how it changes due to deflections of the GEMs \cite{hallermann10}.}
  \label{fig:fieldstack}
\end{figure}

GEMs consist of a thin copper coated insulator foil with microscopic holes etched into it (\autoref{fig:gemmicro}).
Applying a potential difference of a few hundred volts between both copper layers generates high enough field strengths inside of the holes, that gas amplification can take place (\autoref{fig:gemfields}).
This process only depends on the voltage between the copper surfaces.
The efficiencies with which electrons are guided into the holes and with which they are extracted from the holes, collection and extraction efficiency respectively, depend on the respective external field \cite{zenker14}.
These dependencies are roughly linear for moderate changes of the fields around the nominal values for the chosen working conditions, which are shown in the left half of \autoref{fig:fieldstack}.
This means the effective gain in a GEM stack depends not only on the potential differences across the single GEMs, but also on the fields between the GEMs.
Since potentials are constant over the surfaces of the GEMs, deflections of the GEMs change the fields and effective gain as schematically shown in the right half of \autoref{fig:fieldstack}.

The distances between the GEMs and between the last GEM and the readout pad plane (\emph{anode} in \autoref{fig:fieldstack}) in the DESY GridGEM module are \SI{2}{mm} and \SI{3}{mm}, respectively.
The deflections of the GEMs can be in the order of up to a few hundred \si{\micro\meter}.
The effect on the effective gain is therefor expected to be in the order of a few percent.
This makes the effect comparable to the size of the statistical fluctuations of the gas amplification itself.
More importantly, the GEM deflections may introduce a regional bias to the gain so that tracks in different regions would get a different dE$/$dx response.
LEP experience established that this effect should be kept smaller than \SI{10}{\percent} of the desired dE$/$dx resolution, to allow for precise physics analyses relying on dE$/$dx measurements \cite{hallermann10}.

\subsection{Impact on the Point Resolution}
\label{sec:resolution}
\begin{figure}[thp]
  \centering
  \hfill
  \begin{subfigure}{0.45\textwidth}
    \centering
    \includegraphics[width=\textwidth, height=0.25\textheight, keepaspectratio=true]{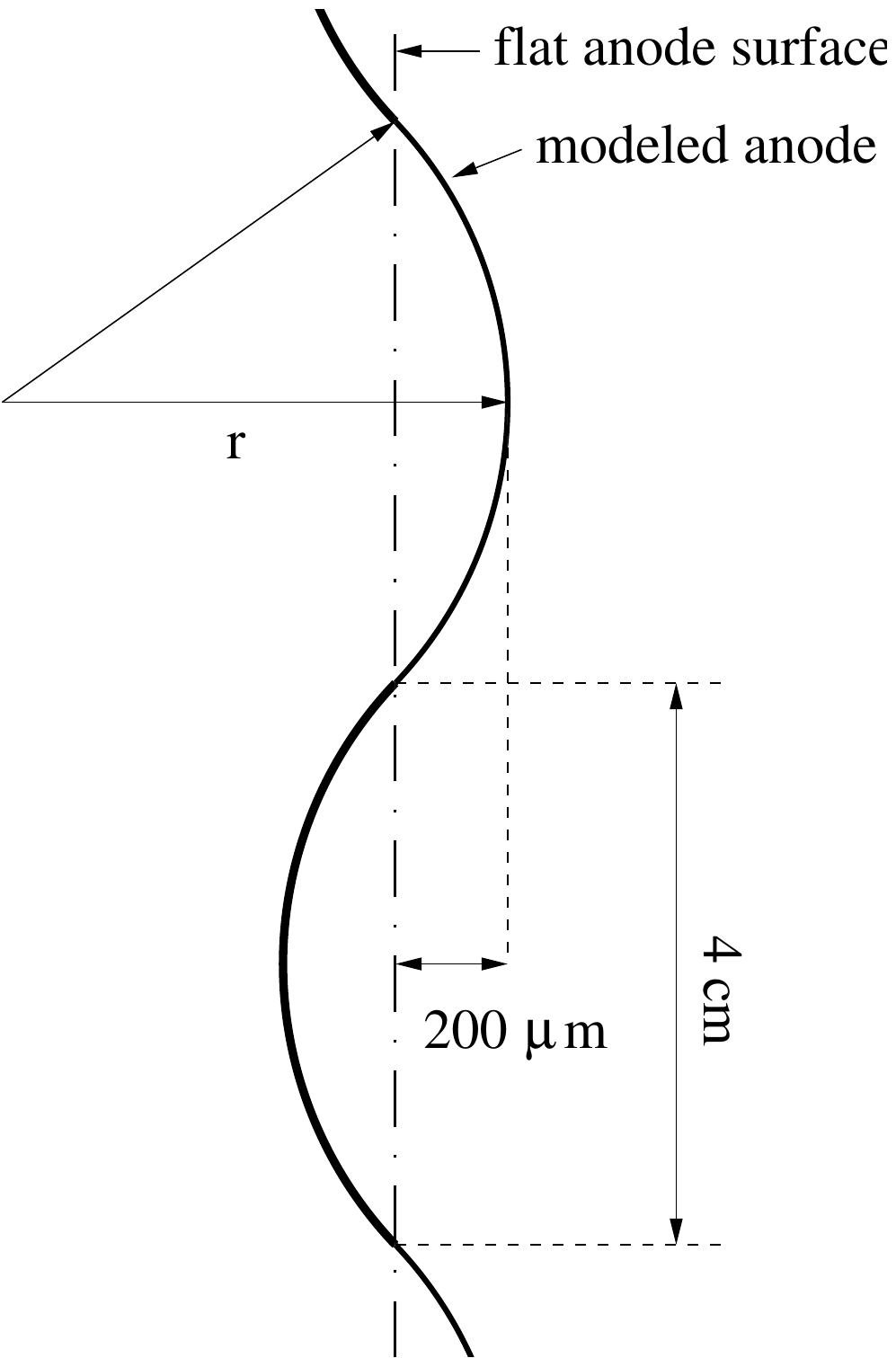}
    \caption{\label{fig:anode_deflections}}
  \end{subfigure}
  \hfill
  \begin{subfigure}{0.45\textwidth}
    \centering
    \includegraphics[width=\textwidth, height=0.25\textheight, keepaspectratio=true]{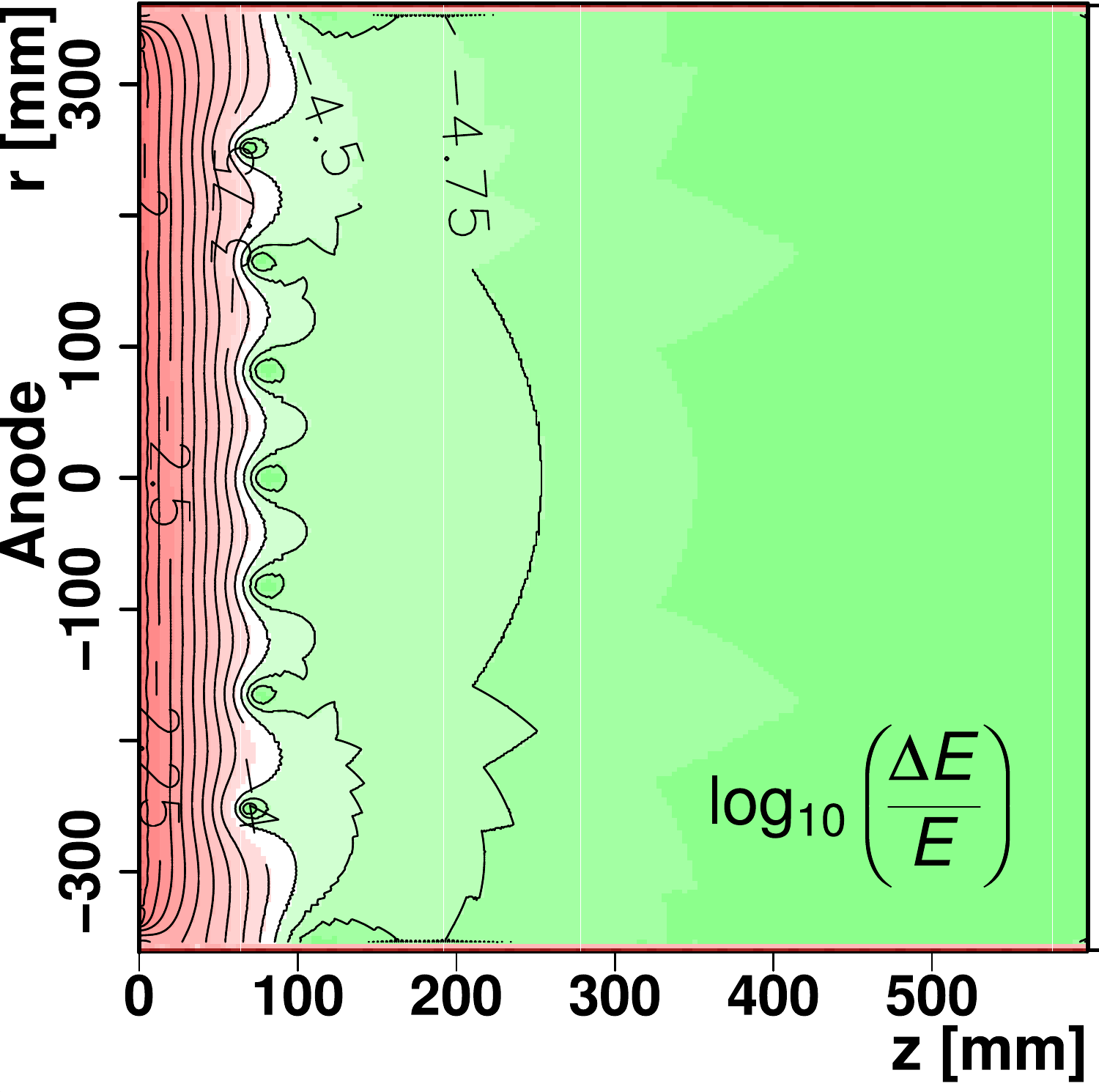}
    \caption{\label{fig:fieldmap}}
  \end{subfigure}
  \hfill
  \caption{\subref*{fig:anode_deflections}) Sketch of the anode shape used in the simulation \cite{hallermann10}. \subref*{fig:fieldmap}) The resulting field quality map. Green areas mark regions, where the required field quality is reached \cite{schade10}.\label{fig:field_simulation}}
\end{figure}

The topmost GEM in a module is one of the electrodes defining the drift field.
This means any deflections of these GEMs introduce inhomogeneities in the drift field.
The effect on the drift field has been simulated assuming a regularly bent anode, seen in \autoref{fig:anode_deflections}, with deflections of \SI{200}{\micro\meter} in both directions and a structure size of \SI{4}{cm}, which are typical dimensions of deflections measured in actual GEMs.
The resulting map of relative field deviations in a large TPC prototype is shown in \autoref{fig:fieldmap}.
It was established that field deviations of up to $\Delta\text{E}/\text{E}=10^{-4}$ are tolerable for the design goal of the ILD TPC regarding single point resolution of $\sigma_{r\phi}<\SI{100}{\micro\meter}$ \cite{hallermann10}.
Depicted in red is an area of about \SI{80}{mm} in front of the anode, where this limit is exceeded.
Close to the anode the simulated distortions are as large as $10^{-2.25}$.

To show the influence on single point resolution, tracks were simulated in the distorted electric field.
The resulting residuals between the reconstructed hits and the actual track position were shown to be smaller than \SI{25}{\micro\meter}, if a \SI{4}{T} magnetic field was assumed.
Assuming these residuals as a systematic uncertainty, the point resolution changes as follows:
\begin{equation*}
\sigma_{r\phi}\rightarrow\sqrt{\sigma_{r\phi}^2+\sigma_{res}^2}=\sqrt{\left(\SI{100}{\micro\meter}\right)^2+\left(\SI{25}{\micro\meter}\right)^2}=\SI{103.1}{\micro\meter}
\end{equation*}
While this effect of about \SI{3}{\percent} still allows reaching the design goal, it shows the importance of keeping the deflections of the anode structure in check.
Especially, since other sources of field distortions, like ion back flow \cite{krautscheid08} add up.

\section{Revising the GEM-Framing Process}
\label{sec:stretchingtool}
\begin{figure}[thp]
  \centering
  \includegraphics[clip=true, trim=0 0 0 5cm, width=0.9\textwidth,height=0.3\textheight,keepaspectratio=true]{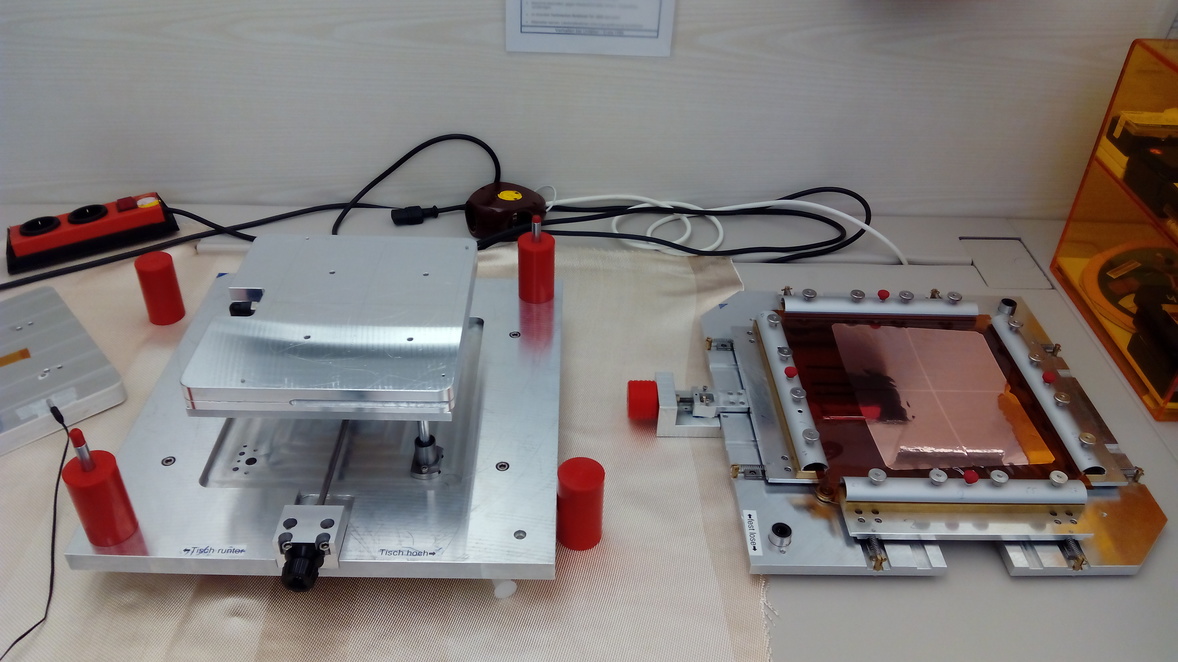}
  \caption{Picture of the tool designed to merge the GEMs and the ceramic frames during gluing. On the left is the base with a lifting stage and a vacuum jig to hold the frame. On the right, the stretching frame holding a GEM is shown.}
  \label{fig:stretchingtool}
\end{figure}

To improve the consistency of the GEM framing process, a tool was developed and commissioned.
This tool consists of a spring loaded stretching frame, which holds the GEM foil and can be separated from the base (\autoref{fig:stretchingtool}).
The tension is relatively low compared to applications using thicker frames.
The base contains a lifting stage, on which a vacuum jig holding the frame can be placed.
The stage serves to accurately merge the ceramic frame and the stretched GEM after glue has been applied to the frame.
The stage is first positioned so, that the ceramic frame does not touch the GEM foil, when the stretching frame is put back onto the base.
Then the stage can be lifted upwards carefully, until ceramic frame and GEM are just touching.
During glue application and merging, the ceramic frames can be held flat by under-pressure, until weight can be applied.
This ensures the glue is applied uniformly to the frame and the frame and GEM are merged consistently everywhere.

\section{Measuring the GEM Flatness}
\label{sec:flatness_measurement}
\begin{figure}[thp]
  \centering
    \includegraphics[clip=true, trim=0 15.5cm 0 10.5cm, width=\textwidth, height=0.3\textheight, keepaspectratio=true]{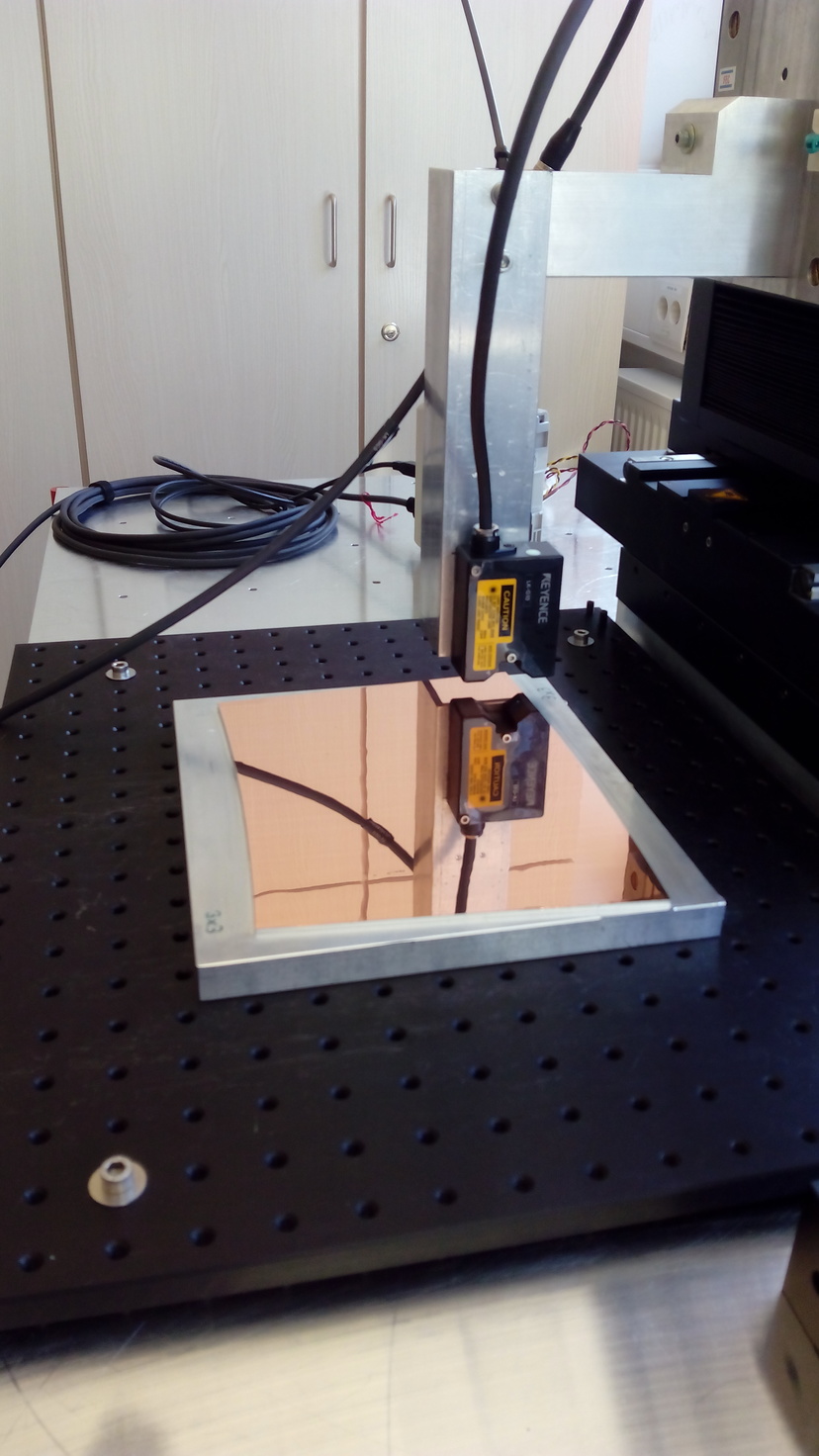}
  \caption{Image of the sensor mounted on the xyz-table measuring a GEM.\label{fig:flatness_setup}}
\end{figure}

A setup to measure height profiles of GEMs was used to evaluate the flatness of the GEMs mounted with the previous method as well as the ones for which the new tool was used.
The measurement setup consists of a laser displacement sensor with a measurement range of \SI{\pm1}{mm} mounted on a precision xyz-movement table as seen in \autoref{fig:flatness_setup}.
The xyz-table is used to meander the sensor over the surface of the GEMs.
It has a positioning accuracy of \SI{3}{\micro\meter} in the axis perpendicular to the GEM surface \cite{lang17}.
The displacement sensor has a measurement repeatability of \SI{0.02}{\micro\meter} \cite{keyence08}.
Both values are much smaller than the expected deflections in the GEMs.

\subsection{Evaluating the new Framing Process}
\label{sec:flatness_results}
\begin{figure}[thp]
  \centering
  \hfill
  \begin{subfigure}{0.45\textwidth}
    \centering
    \includegraphics[width=\textwidth, height=0.3\textheight, keepaspectratio=true]{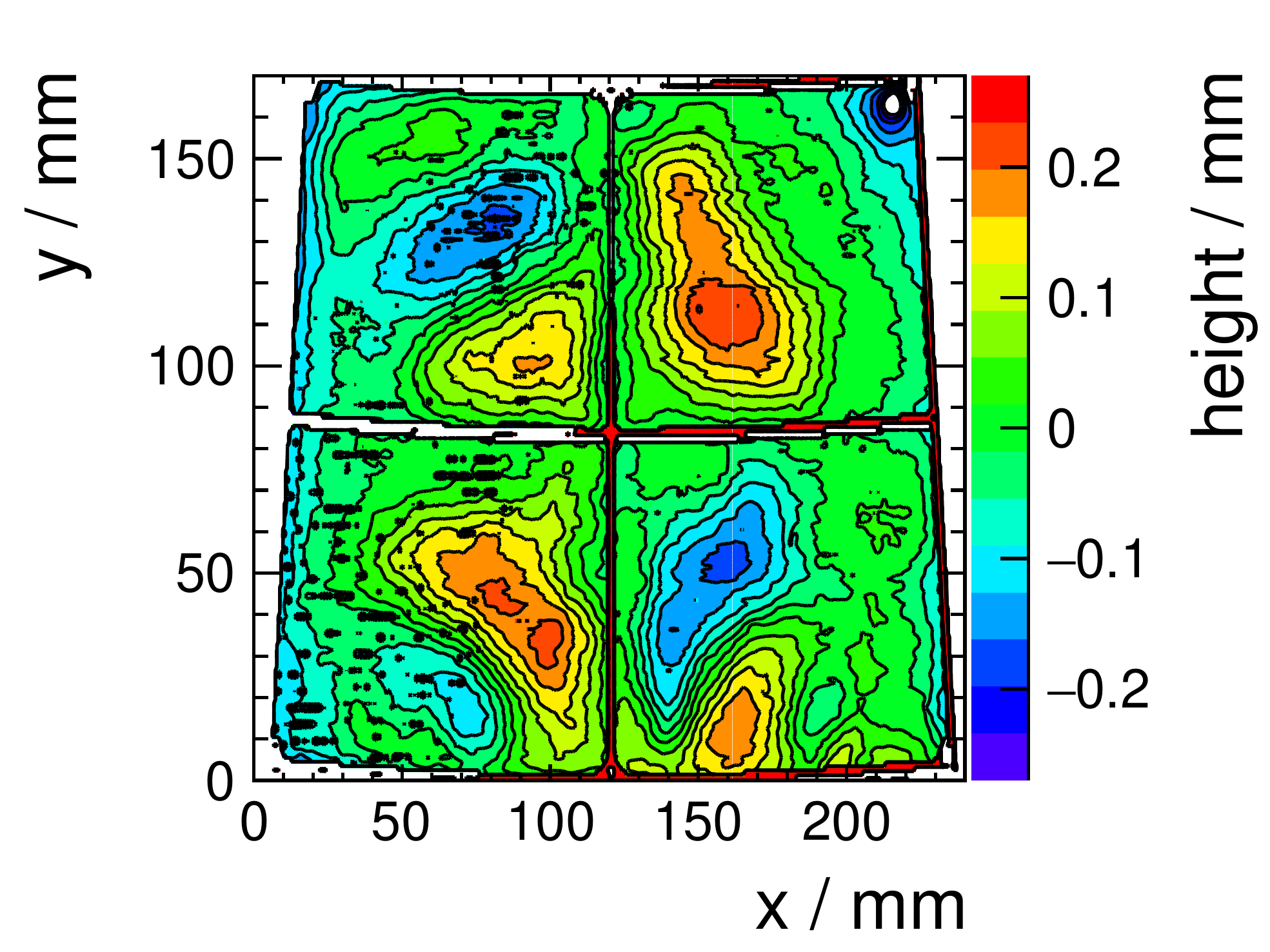}
    \caption{Last generation GEM\label{fig:gem_profile_old}}
  \end{subfigure}
  \hfill
  \begin{subfigure}{0.45\textwidth}
    \centering
    \includegraphics[width=\textwidth, height=0.3\textheight, keepaspectratio=true]{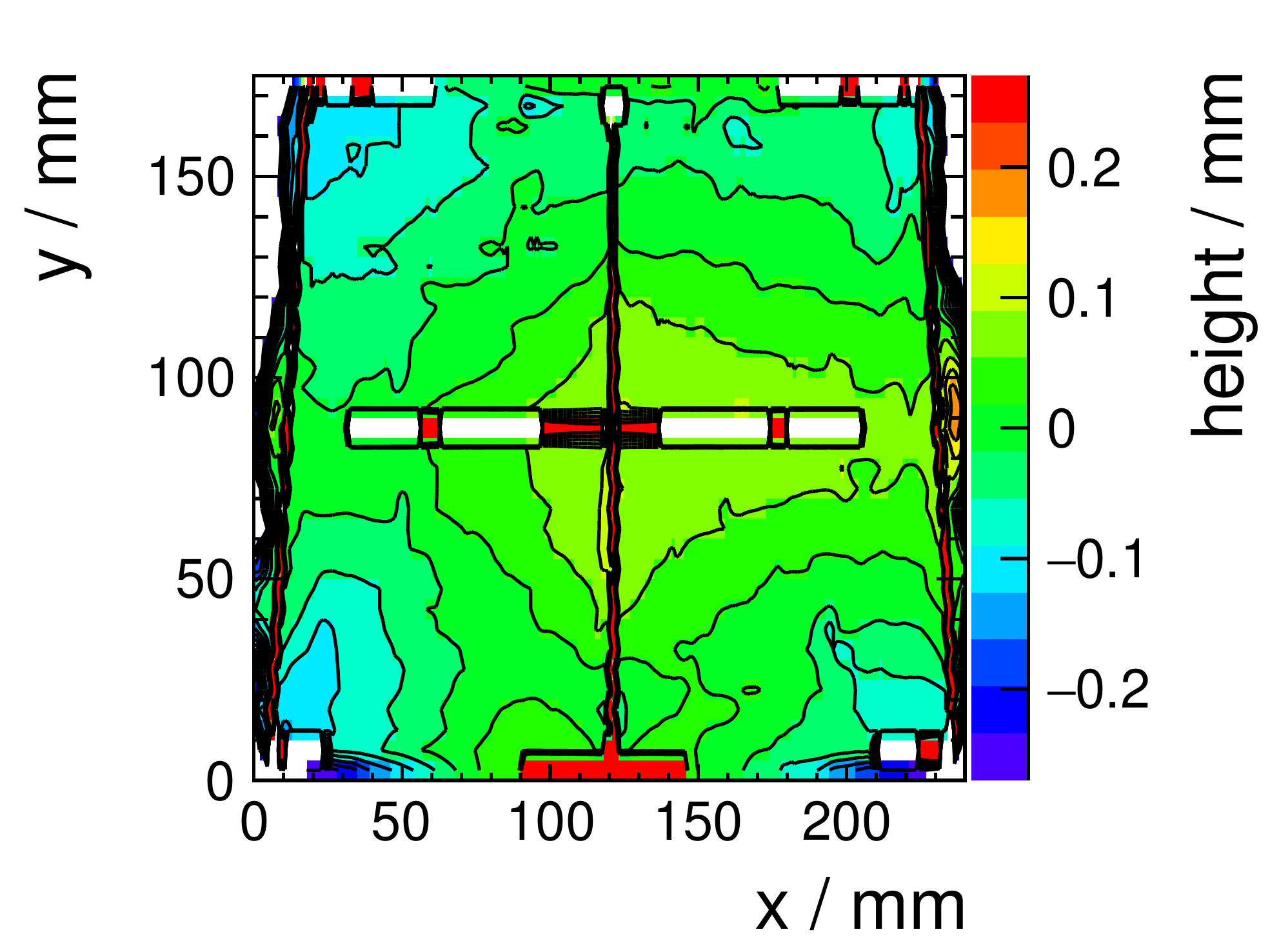}
    \caption{Current generation GEM\label{fig:gem_profile_new}}
  \end{subfigure}
  \hfill
  \caption{Two examples of measured GEM height profiles mounted \subref*{fig:gem_profile_old}) without using the new tool and \subref*{fig:gem_profile_new}) using it. In both cases the ceramic frame on top of the GEM is partly visible as red areas, where it was just in the measurement range of the sensor. \label{fig:gem_profiles}}
\end{figure}

In \autoref{fig:gem_profiles}, examples for height profiles of an old and a new GEM are shown.
For the last generation GEM (\autoref{fig:gem_profile_old}) the depressions and elevations of the GEM sectors in the order of \SI{200}{\micro\meter} are clearly visible as blue and red areas, respectively.
The example of the new GEM in \autoref{fig:gem_profile_new} provides a much better result, as there are only minor deflections visible in the four sectors.
Here, the deflections seem to mostly come from the curvature of the frame.

\begin{figure}[thp]
  \centering
  \hfill
  \begin{subfigure}{0.45\textwidth}
    \centering
    \includegraphics[width=\textwidth, height=0.25\textheight, keepaspectratio=true]{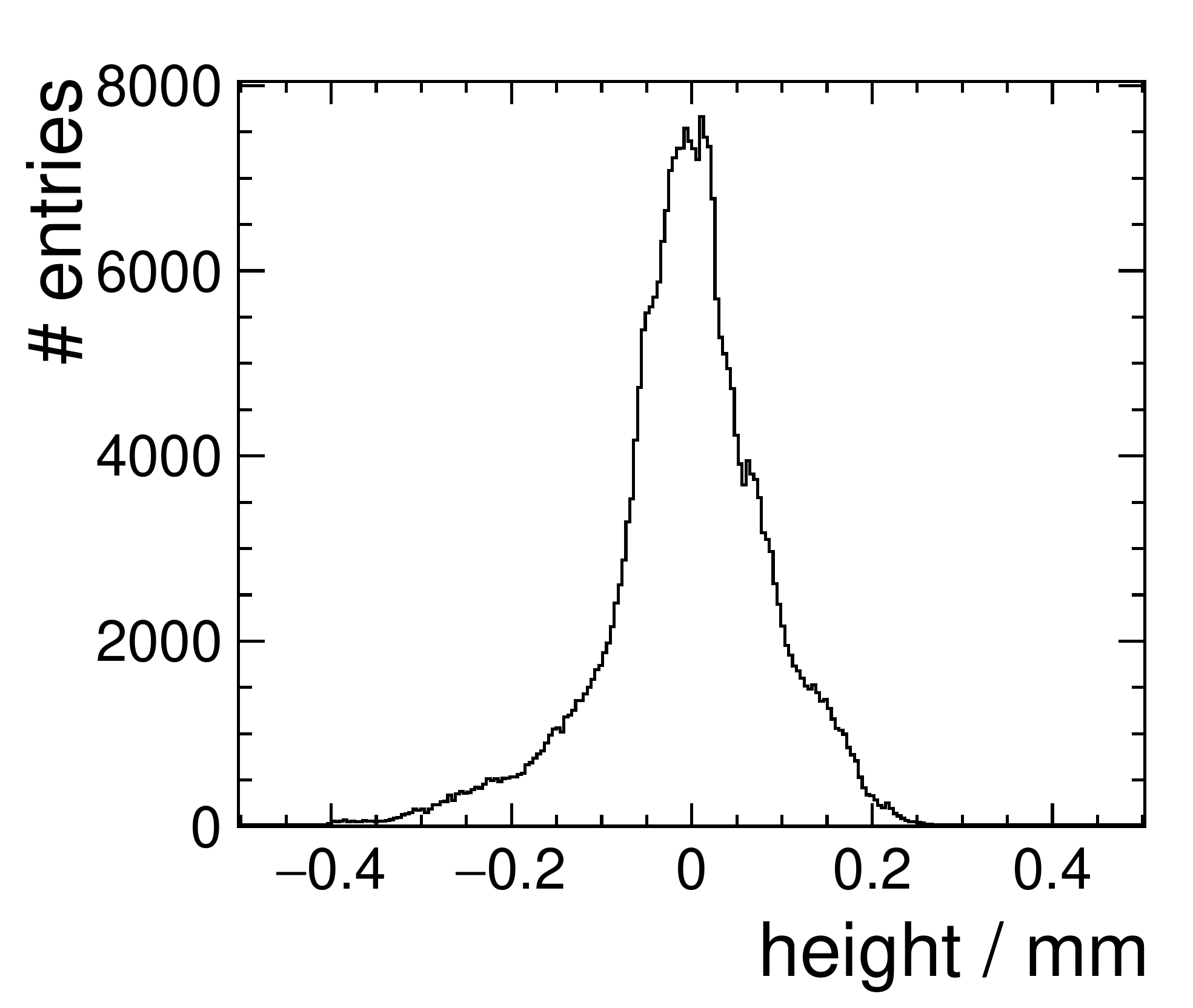}
    \caption{Last generation GEMs\label{fig:height_distributions_old}}
  \end{subfigure}
  \hfill
  \begin{subfigure}{0.45\textwidth}
    \centering
    \includegraphics[width=\textwidth, height=0.25\textheight, keepaspectratio=true]{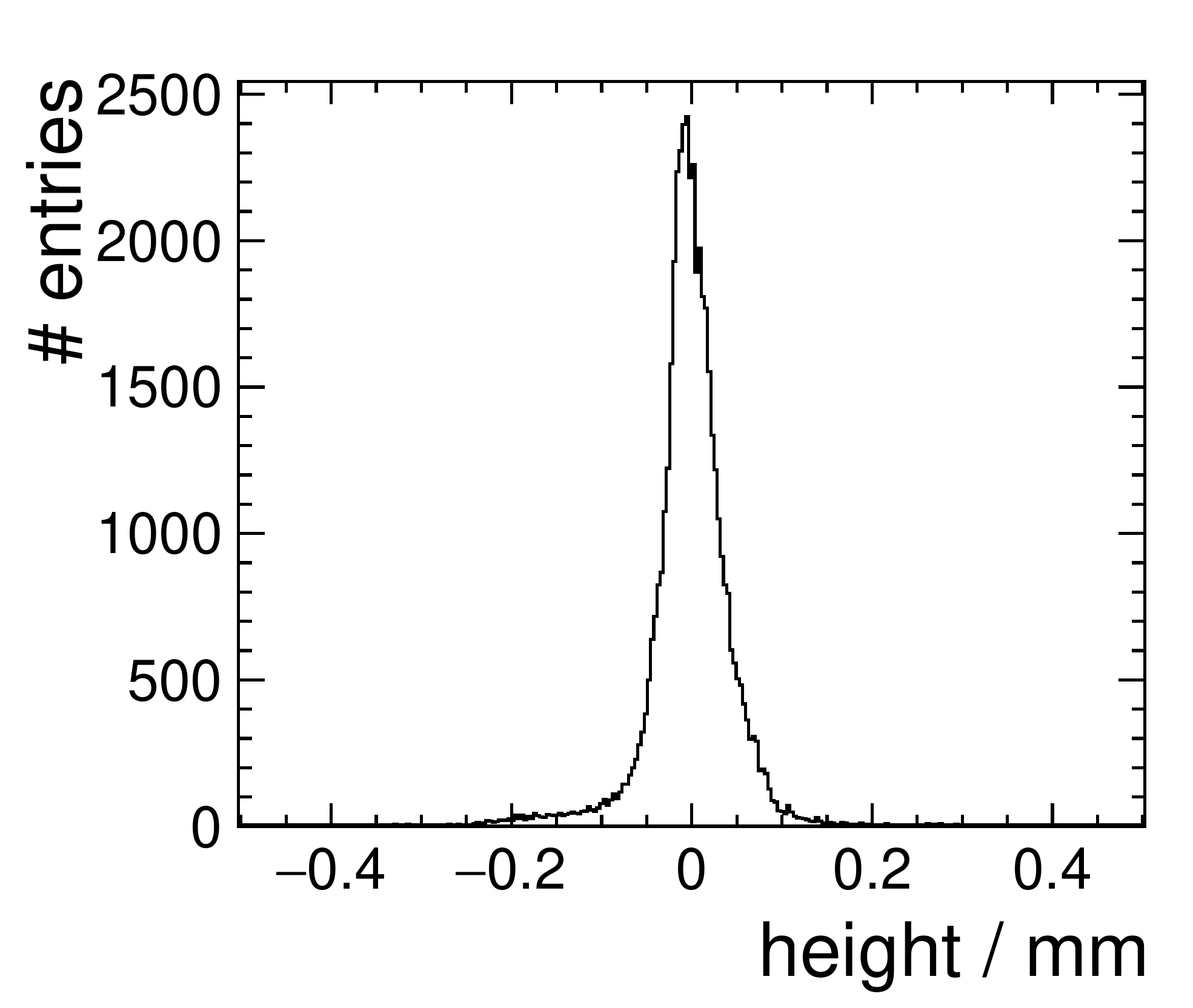}
    \caption{Current generation GEMs\label{fig:height_distributions_new}}
  \end{subfigure}
  \hfill
  \caption{Combined height distributions of all measured GEMs: \subref*{fig:height_distributions_old}) for those framed without using the new tool; \subref*{fig:height_distributions_new}) for the ones framed using it.\label{fig:height_distributions}}
\end{figure}

\autoref{fig:height_distributions_old} and \subref{fig:height_distributions_new} show the combined height distributions for all measured old and new GEMs, respectively.
These distributions contain tails from measurement points on the ceramic frames and on excess foil outside of the frame, which was not cut away for some GEMs (visible in \autoref{fig:gem_profile_new}).
To reduce the bias introduced by these points, the RMS of the central \SI{95}{\percent} of the distributions is used to compare the two sets.
The measured last generation GEMs show a width of the height distribution of \SI{68}{\micro\meter} and the width of the single GEM distributions spread from \SI{50}{\micro\meter} to \SI{90}{\micro\meter}.
The combined height distribution of the new GEMs has a width of \SI{35}{\micro\meter} and the single GEMs show values from \SI{20}{\micro\meter} to \SI{50}{\micro\meter}.
This shows, that by using the new assembly tool, the deflections of the GEMs could consistently be reduced by almost a factor of two.

\subsection{Impact on the effective Gas Gain}
\label{sec:gain}
An existing parametrization of the effective gain in a triple GEM stack \cite{lotze06} is utilized to evaluate the influence of the GEM deflections.
Input to this parametrization are the voltages across the GEMs and the fields between them, as well as gas and magnetic field parameters.
Here parameters for T2K-TPC gas without magnetic field are used.
T2K gas consists of \SI{95}{\percent} Argon, \SI{3}{\percent} $\text{CF}_4$ and \SI{2}{\percent} isobutane \cite{t2k11}.
The GEM height profiles serve as input to recalculate the transfer and induction fields between GEMs and between the last GEM and the readout plane, respectively.
This is done for each geometrical bin of the height profiles in the active area of the module and for each possible combination of GEMs.
Since the number of measured GEMs is limited, the statistics is improved by mirroring each profile at its symmetry axis and in addition by inverting its height map.
Stacks containing only mirrored profiles were ignored to avoid double counting.
Also stacks containing a profile and its inverted version were not taken into account, because the effects would be exaggerated.

\begin{figure}[thp]
  \centering
  \hfill
  \begin{subfigure}{0.45\textwidth}
    \centering
    \includegraphics[width=\textwidth, height=0.25\textheight, keepaspectratio=true]{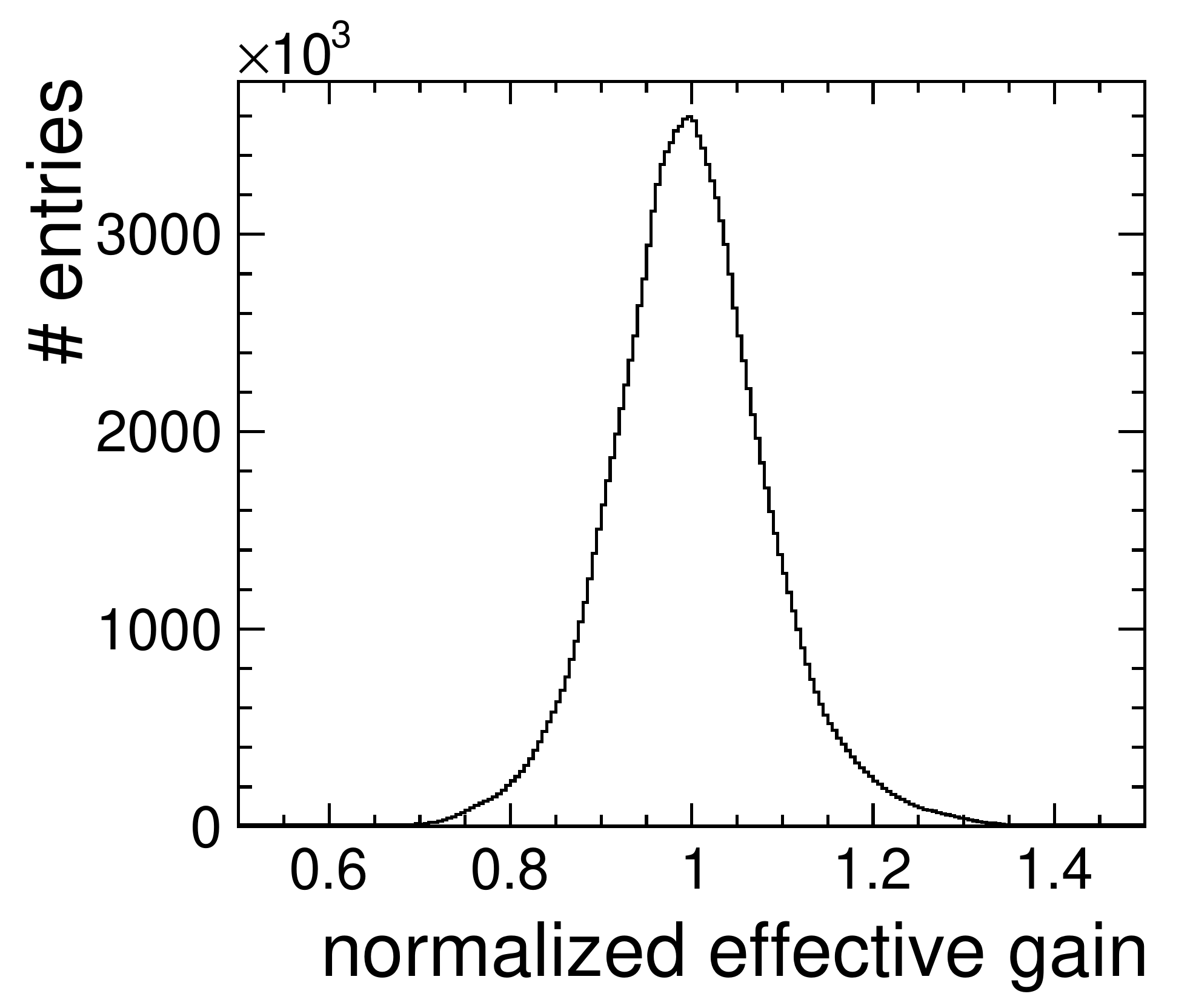}
    \caption{Last generation GEMs\label{fig:gain_dist_old}}
  \end{subfigure}
  \hfill
  \begin{subfigure}{0.45\textwidth}
    \centering
    \includegraphics[width=\textwidth, height=0.25\textheight, keepaspectratio=true]{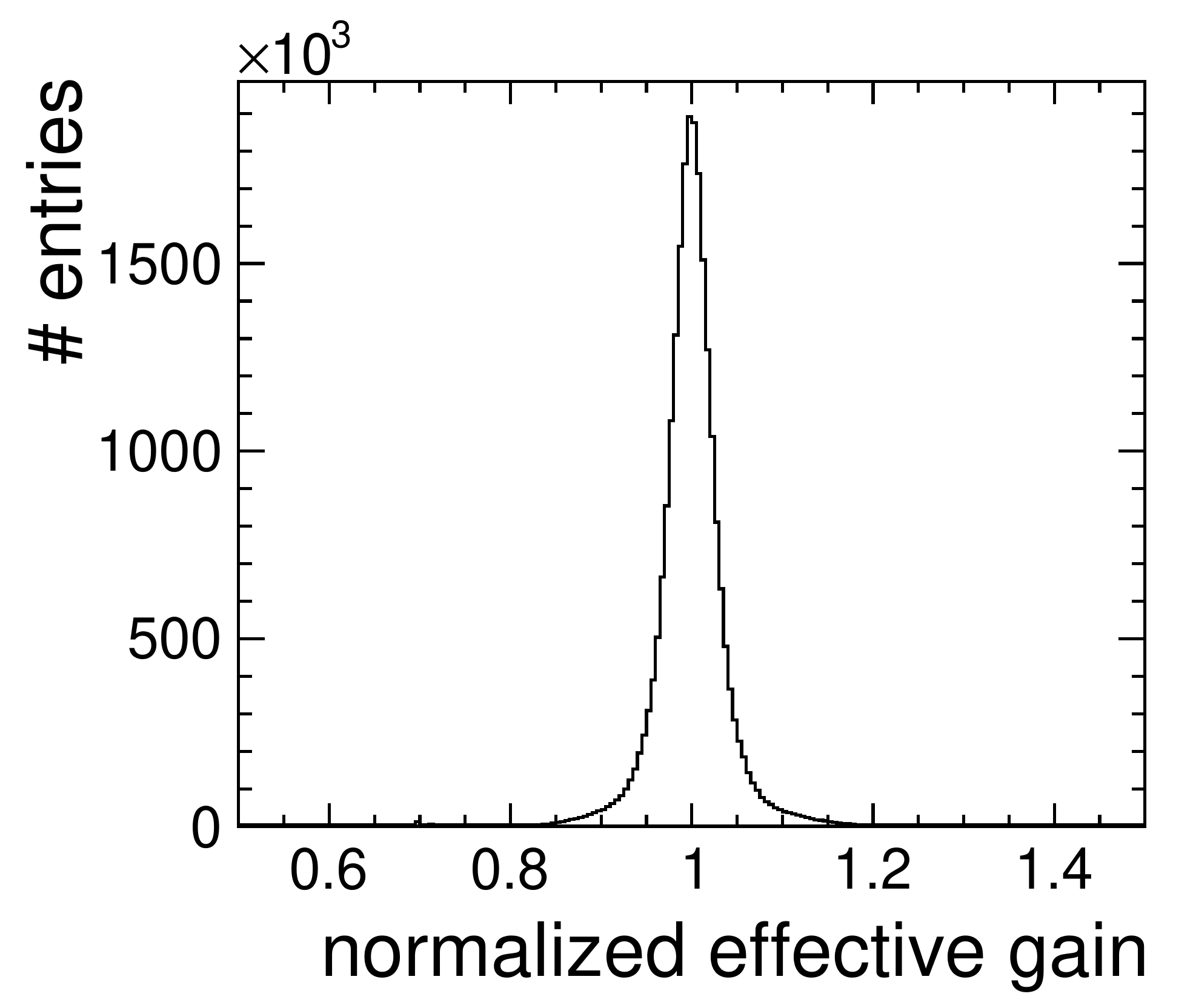}
    \caption{New GEMs\label{fig:gain_dist_new}}
  \end{subfigure}
  \hfill
  \caption{Combined distributions of the effective gain normalized to the nominal gain for stacks build out of GEMs framed \subref*{fig:gain_dist_old}) without using the new tool ($\text{RMS}=\SI{6.1}{\percent}$) and \subref*{fig:gain_dist_new}) for the ones framed using it ($\text{RMS}=\SI{4.2}{\percent}$).\label{fig:gain_distributions}}
\end{figure}
The resulting gain distributions, normalized to the expected gain, for old and new GEMs can be seen in \autoref{fig:gain_dist_old} and \subref{fig:gain_dist_new}, respectively.
The width of those distributions can be directly translated into a systematic uncertainty on the dE$/$dx measurement as described in \cite{hallermann10}.
For the last generation GEMs this uncertainty would be \SI{6.1}{\percent}.
With the new GEM mounting method this could be reduced to \SI{4.2}{\percent}.




\section{Summary}
\label{sec:summary}
The last DESY GridGEM module generation was investigated and several points for possible improvements were identified.
For the reasons given in \autoref{sec:performance}, the focus was set on improving the flatness of the framed GEMs.
A new tool was designed to aid in the gluing and framing process.
Using this, the average deflections of the GEM foils could consistently be reduced by a factor of two.
This improved the gain homogeneity from \SI{6.1}{\percent} to \SI{4.2}{\percent}

\printbibliography
\end{document}